\documentclass[aps,pra,
    twocolumn,
	superscriptaddress,
	showpacs,preprintnumbers,amsmath,amssymb,nofootinbib,
	longbibliography
	]{revtex4-2}
\PassOptionsToPackage{dvipsnames}{xcolor}

\usepackage[
	colorlinks=true,
	pdfstartview=FitV,
	linkcolor=blue,
	citecolor=magenta,
	urlcolor=blue,
	bookmarks=true,
	bookmarksnumbered=true,
	pdftitle={},
	pdfauthor={}
]{hyperref}

\usepackage{bbm,bm,mathtools,color,slashed}
\usepackage{amssymb,amsmath}
\usepackage[T1]{fontenc}

\usepackage[all]{xy}

\usepackage{graphicx}

\usepackage[framemethod=tikz]{mdframed}
\usepackage{tikz}

\usepackage[normalem]{ulem}  % \sout{old text} for strikeout

\usepackage[most]{tcolorbox}

\newcommand{\cout}[1]{ \if 0 {#1} \fi }

\newcommand{\diff}{\mathrm{d}} 
\newcommand{\rmi}{\mathrm{i}} 
\newcommand{\rme}{\mathrm{e}}

\newcommand{\hH}{\hat{H}}

\newcommand{\tilvarphi}{\widetilde{\varphi}}
\newcommand{\tilalpha}{\widetilde{\alpha}}

\newcommand{\ptc}[1]{{\bar{#1}}}

\newcommand{\hb}{\hat{b}}

\newcommand{\hpi}{\hat{\pi}}

\newcommand{\halpha}{\hat{\alpha}}

\newcommand{\pn}{\ptc{n}}
\newcommand{\pp}{\ptc{p}}

\newcommand\Lcal{\mathcal{L}}

\newcommand\Acal{\mathcal{A}}

\newcommand\Vcal{\mathcal{V}}

\newcommand{\ket}[1]{| {#1} \rangle}

\newcommand{\bx}{\bm{x}}

\newcommand{\bnab}{\bm{\nabla}}

\newcommand{\U}{\text{U}}

\newcommand{\const}{\mathrm{const.}}
\newcommand{\eff}{\mathrm{eff}}

\begin{document}
%\preprint{}

\title{
 Effective theory of surface oscillations in self-bound superfluid droplets
}

\author{Jun Mitsuhashi}
\email{f25a016j@mail.cc.niigata-u.ac.jp}
\affiliation{Graduate School of Science and Technology, Niigata University,  Niigata, 950-2181, Japan}

\author{Keisuke Fujii}
\email{fujii@phys.sci.isct.ac.jp}
\affiliation{Department of Physics, Institute of Science Tokyo,
Ookayama, Meguro-ku, Tokyo 152-8551, Japan}

\author{Masaru Hongo}
\email{hongo@phys.sc.niigata-u.ac.jp}
\affiliation{Department of Physics, Niigata University, Niigata 950-2181, Japan}
\affiliation{RIKEN iTHEMS, RIKEN, Wako 351-0198, Japan}

\date{\today}

\begin{abstract}

We investigate the low-energy dynamics of small-amplitude surface oscillations of spherical superfluid droplets in vacuum.
Starting from the effective field theory of superfluid phonons, we derive an effective action governing the surface oscillations under a fixed particle-number constraint.
The normal-mode eigenfrequencies $\omega_{\ell}$ for each angular momentum quantum number $\ell$ are determined and shown to depend on a dimensionless parameter measuring the ratio of surface tension to bulk compressibility energy.
We identify a critical value of this parameters at which the breathing mode ($\ell = 0$) becomes mechanically unstable, and show that all multipole surface modes with $\ell \geq 2$ enter the low-energy regime when the surface tension is sufficiently small.
Within this regime, we further quantize the surface oscillations, whose quanta correspond to ripplons, allowing the construction of general multi-ripplon states obeying angular-momentum selection rules.
We also apply our formalism to a concrete example: a weakly interacting two-component Bose mixture realizing a self-bound superfluid droplet.
The resulting description is universal in the sense that it applies to surface dynamics of generic nonrelativistic superfluids with a free interface, independent of microscopic details.
\end{abstract}

\maketitle

\section{Introduction}

Dynamics associated with fluid interfaces is a long-standing problem in both classical and quantum physics.
The analysis of surface dynamics in classical fluids dates back to the late nineteenth century.
Canonical results include propagating capillary waves localized near a flat, extended interface, established by William Thomson (Lord Kelvin)~\cite{Thomson1871}, as well as surface oscillations of an incompressible spherical droplet derived by Lord Rayleigh~\cite{Rayleigh1879}.
At the heart of these phenomena lies the surface tension, which serves as a restoring force for a fluid interface and gives rise to characteristic surface-wave modes (see, e.g., Refs.~\cite{Landau:Statistical,Landau:Fluid}).
How such surface-tension–driven modes are described and quantized in finite quantum fluids remains a nontrivial issue.

In quantum systems, surface excitations of finite droplets have been investigated in several distinct physical contexts.
In nuclear physics, the surface dynamics of finite quantum droplets has been extensively studied within the liquid-drop description of atomic nuclei.
In this context, the quantization of small-amplitude surface oscillations accounts for collective excitation spectra, while large deformations provide a theoretical basis for nuclear fission~\cite{Bohr:1939ej}.
Furthermore, the liquid-drop model, when unified with single-particle excitations, plays a crucial role in understanding various properties of deformed nuclei~\cite{Rainwater1950,Bohr1952,Bohr-Mottelson1953,Bohr-Mottelson1974}.
In these approaches, however, the bulk low-energy degrees of freedom associated with spontaneous symmetry breaking due to Bardeen--Cooper--Schrieffer-type pairing of nucleons~\cite{Bohr-Mottelson-Pines1958,Belyaev1959}---namely, the phase mode (see also e.g.,~\cite{RingSchuck1980,RevModPhys.75.607,BrogliaZelevinsky2013} for reviews), which becomes a gapless superfluid phonon in the thermodynamic limit---are typically not treated as explicit dynamical modes.

In contrast, collective modes of superfluids have long been studied in various contexts. 
In particular, superfluid helium droplets have served as paradigmatic examples of self-bound quantum fluids, whose collective excitations have been extensively investigated both experimentally and theoretically~\cite{Casas:1990,Barranco:2006}. 
For atomic Bose-Einstein condensates (BECs), collective modes such as breathing oscillations and surface deformations have been widely analyzed within hydrodynamic descriptions and the Gross-Pitaevskii framework, in which the gapless superfluid phonon associated with spontaneous U(1) symmetry breaking is explicitly incorporated.
Until recently, studies of BEC collective excitations have focused mainly on ultracold atomic gases confined in trapping potentials.
More recently, following the theoretical proposal of self-bound quantum droplets in Bose mixtures~\cite{Petrov:2015,Petrov:2016}, experimental and theoretical exploration of quantum droplets in free space has become increasingly active~\cite{Barbut:2019,Malomed:2020,Bottcher:2020,Reimann:2023}.
Quantum droplets have indeed been observed
in spin mixtures of $^{39}$K~\cite{Cabrera:2018,Semeghini:2018},
in heteronuclear mixtures of $^{41}$K-$^{87}$Rb~\cite{DErrico:2019,Burchianti:2020,Cavicchioli:2025} and $^{23}$Na-$^{87}$Rb~\cite{Guo:2021}.

These experimental realizations have, in turn, facilitated detailed studies of collective excitations and their spectra in self-bound droplets.
For nearly spherical droplets, surface-localized modes have been identified as part of the density and phase fluctuation spectra~\cite{Hui:2020}.
In particular, surface waves, so-called ripplons, exhibit a fractional-power dispersion relation, $\omega\sim k^{3/2}$ at low wavenumber in the flat-surface limit.
This non-analytic behavior originates from the gapless nature of superfluid phonons propagating in the bulk and indicates that low-energy bulk degrees of freedom fundamentally govern the surface dynamics.
For an infinitely extended flat superfluid interface, this fractional dispersion relation has been elucidated within an effective field theory (EFT) framework~\cite{Watanabe:2014zza}.
In that EFT framework, both the bulk phonons and the surface are treated as explicit low-energy dynamical degrees of freedom, allowing for a transparent derivation of the non-analytic low-energy behavior dictated by the gapless bulk modes.

By contrast, for finite self-bound superfluid droplets with spherical geometry, the situation is richer.
In addition to the bulk and surface energy contributions characterizing a finite spherical droplet, global particle-number conservation imposes an additional constraint, leading to a more intricate coupling structure between surface modes and bulk phonons.
While most quantum droplets realized to date are multi-component mixtures and support an additional gapless spin (out-of-phase) mode besides the density (in-phase) mode, surface deformations predominantly couple to fluctuations of the total density profile.
Accordingly, we focus on the low-energy sector relevant for surface dynamics, where interface fluctuations couple to the gapless density phonon, and adopt the simplest effectively single-component setting as a minimal framework capturing the essential bulk-surface interplay.

Motivated by these considerations, in this work we develop an EFT framework for surface oscillations of finite superfluid droplets that explicitly incorporates the surface as a dynamical degree of freedom.
Building on the EFT of superfluid phonons, we demonstrate how gapless bulk phonon excitations govern the low-energy dynamics of the droplet surface.
Our approach provides an effective description of surface modes in finite spherical geometries and naturally reproduces the characteristic non-analytic behavior known from flat superfluid interfaces in the appropriate limit.
In this way, we establish a universal low-energy description of surface oscillations in finite self-bound superfluid droplets based on the EFT of superfluid phonons.

The organization of this paper is as follows.
In Sec.~\ref{sec:model}, we introduce the effective field theory of superfluids with a free boundary as our starting point.
In Sec.~\ref{sec:fluctuation}, we consider small fluctuations around the spherical background and derive the quadratic action for the whole system.
In Sec.~\ref{sec:vacuum}, we derive an effective theory of surface oscillations by integrating out bulk superfluid phonons, present the energy spectrum of the surface modes for each angular momentum, and discuss their quantization in the low-energy regime, leading to ripplons.
In Sec.~\ref{sec:application-cold-atom}, we discuss an application of the formalism to a cold atomic system, namely a weakly interacting two-component Bose gas.
Finally, Sec.~\ref{sec:summary} contains concluding remarks.
Throughout this paper, we use units where $\hbar = 1$.

\section{Model}
\label{sec:model}

In this section, we present an effective field theory for a self-bound superfluid droplet in vacuum.
In Sec.~\ref{sec:Lagrangian}, we introduce the effective Lagrangian for a superfluid with a free boundary.
After specifying the conserved current and the particle-number constraint appropriate for an isolated droplet in Sec.~\ref{sec:constraint}, we derive the equations of motion and boundary conditions in Sec.~\ref{sec:eom}.

\subsection{Superfluid EFT with a free boundary}
\label{sec:Lagrangian}

We consider a self-bound, nearly spherical superfluid droplet in vacuum.
The system possesses a global $\mathrm{U}(1)$ symmetry associated with conservation of the total particle number, which is spontaneously broken in the superfluid phase.
The droplet is isolated from any particle reservoir, so that its total particle number is fixed.
We further assume that the droplet is sufficiently large compared to the healing length near the surface, allowing the interior to be treated as approximately uniform.
Under these assumptions, the system can be modeled as a uniform superfluid droplet separated from the vacuum by a well-defined surface.

\begin{figure}
    \centering
    \includegraphics[trim ={2cm 2.3cm 0.3cm 2cm} ,clip, width=0.7
\linewidth]{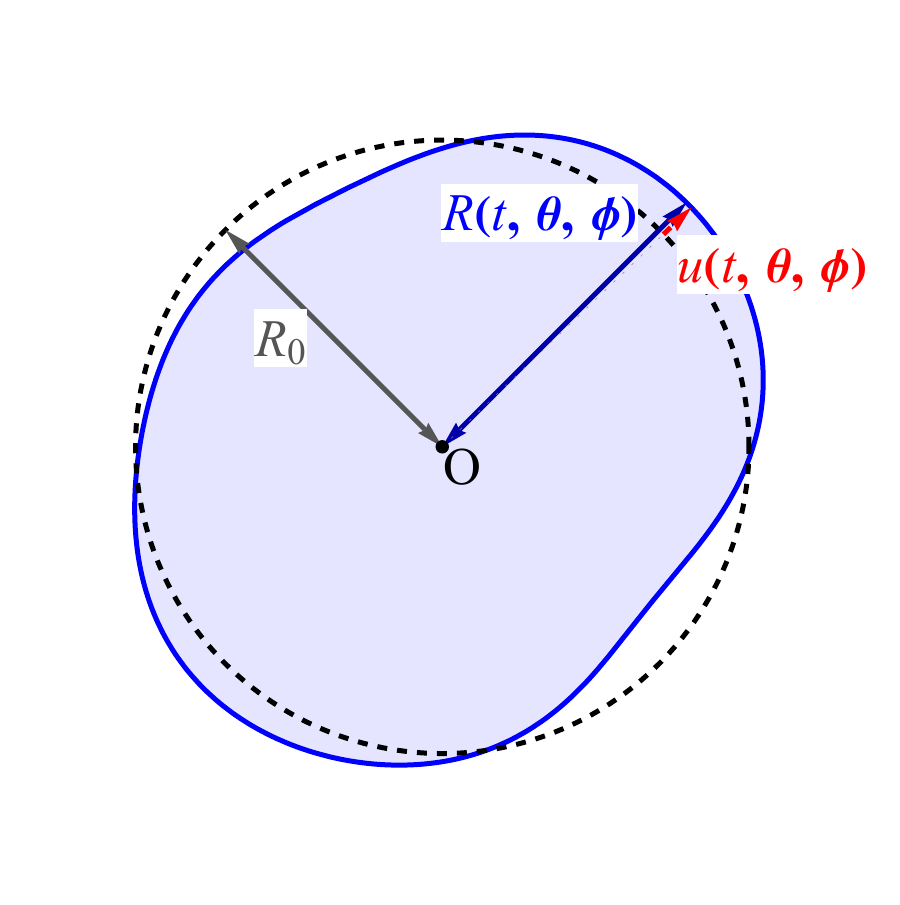}
    \caption{
    Schematic cross-sectional view of the droplet, illustrating the time-dependent radius $R(t,\theta,\phi)$, the equilibrium radius $R_0$, and the radial displacement field $u(t,\theta,\phi)$.
    }
   \label{fig:R-u-demonstration}
\end{figure}

We suppose that the ground-state shape of the droplet is spherical with radius $R_0$.
We introduce a spherical coordinate system $x^i = (r,\theta,\phi)$ whose origin is located at the center of the droplet.
Small-amplitude fluctuations around the spherical ground state at time $t$ are described by a radial displacement field $u(t,\theta,\phi)$.
We then parametrize the position of the interface as
\begin{equation}
 R(t,\theta,\phi) 
 = R_0 + u(t,\theta,\phi),
 \label{eq:radius-R}
\end{equation}
as illustrated in Fig.~\ref{fig:R-u-demonstration}.

We now note that the superfluid occupies the inner region $r<R$, while the outer region $r>R$ is vacuum, where the interface radius $R$ is defined in Eq.~\eqref{eq:radius-R}.
We treat the interface radius $R$ as a time-dependent dynamical variable coupled to the low-energy degrees of freedom in the bulk superfluid.
The crucial point is that the dependence on $R$ enters in two distinct ways: first, through the specification of the spatial region occupied by the bulk superfluid, and second, through the surface-tension term.
Under the above assumptions, the total effective Lagrangian consists of a bulk superfluid contribution together with a surface-tension term.

Due to the spontaneous breaking of the global $\U(1)$ symmetry, the low-energy dynamics of the superfluid is governed by the associated Nambu–Goldstone mode, namely the superfluid phonon.
To describe the bulk superfluid, we employ an effective field theory for superfluid phonons, whose Lagrangian density is given by~\cite{Greiter:1989,Son:2005rv}
\begin{equation}
 \mathcal{L}_{\varphi}
 = p (X)
 ~~\mathrm{with}~~
 X := \mu - \partial_t \varphi
 - \frac{(\bnab\varphi)^2}{2M},
 \label{eq:phonon-Lagrangian}
\end{equation}
where $\mu$ denotes the chemical potential, $M$ is the mass of the microscopic constituent, and $\varphi=\varphi(t,\bx)$ is the superfluid phonon field.
As we will see shortly, the chemical potential $\mu(t)$ is allowed to be time dependent in order to enforce the constraint on the conservation of the total particle number.
The functional form of $p(X)$ is determined by the equation of state, namely the pressure $p(\mu)$ as a function of the chemical potential.

It is worth emphasizing that the effective Lagrangian~\eqref{eq:phonon-Lagrangian} provides a universal low-energy description of a superfluid, with microscopic details entering only through the equation of state.
We note, however, that the existence of a self-bound droplet with a thin interface restricts the class of microscopic models that can realize a self-bound phase.
A systematic investigation of these constraints for specific microscopic theories is an interesting problem in its own right, but lies beyond the scope of the present work.
As a concrete example, we will later consider in Sec.~\ref{sec:application-cold-atom} a weakly interacting two-component Bose gas.

With these ingredients, the effective action $S[\varphi,R,\mu]$ for the entire system is given by
\begin{widetext}
\begin{align}
 S
 &= \int_{r<R} \diff t \diff^3 x\, \sqrt{g} p (X)
  - \sigma \int \diff t\, \Acal [R]
  - \int \diff t\, \mu (t) N
 \label{eq:action}
\end{align}
where the integration domain depends on time and the angular coordinates through $R = R(t,\theta,\phi)$.
Here, to implement the particle-number constraint, we introduce an additional term $\int dt\, \mu(t) N$, where $N$ is a fixed constant.
Together with the $\mu(t)$-dependent contribution already contained in the first term of Eq.~\eqref{eq:action}, this construction promotes the chemical potential $\mu(t)$ to a time-dependent Lagrange multiplier.
In this way, the total particle number is fixed through the variational principle.

The second term of Eq.~\eqref{eq:action} represents the surface-tension contribution proportional to the surface tension $\sigma$ and the surface area $\Acal[R]$.
For the radius $R(t,\theta,\phi)$, the surface area $\Acal[R]$ is given by
\begin{equation}
 \Acal[R]
 :=
 \int \diff \Omega\,
 R^2 
 \sqrt{ 1 + (\bnab R)^2} 
 =
 \int \diff \Omega\,
 R^2 
 \sqrt{
 1
 + \frac{1}{R^2}
 \left[
 (\partial_\theta R)^2
 + \frac{1}{\sin^2\theta}(\partial_\phi R)^2
 \right]} 
 ,
\end{equation}
\end{widetext}
with the solid-angle measure $\diff \Omega := \diff \theta \diff \phi \sin \theta$.
The surface area reduces to $\mathcal{A}[R_0]=4\pi R_0^2$ for a spherical ground-state configuration $R(t,\theta,\phi)=R_0$.
We introduce the notation $\partial_i=(\partial_r,\partial_{\theta},\partial_{\phi})$ together with the metric $g_{ij}$ and its inverse $g^{ij}$,
\begin{equation}
  g_{ij} =
  \begin{pmatrix}
      1 & 0 & 0 \\
      0 & r^2 & 0 \\
      0 & 0 & r^2 \sin^2 \theta
  \end{pmatrix},
  ~~
  g^{ij} =
  \begin{pmatrix}
      1 & 0 & 0 \\
      0 & \frac{1}{r^2} & 0 \\
      0 & 0 & \frac{1}{r^2 \sin^2\theta}
  \end{pmatrix},
  \label{eq:metric}
\end{equation}
which yields the Jacobian $\sqrt{g} = \sqrt{\det g_{ij}} = r^2 \sin \theta$.
With this convention, $X$ is expressed as $X = \mu - \partial_t \varphi - g^{ij} \partial_i \varphi \partial_j \varphi/(2M)$, where repeated spatial indices $i,j,\ldots$ are implicitly summed over $(r,\theta,\phi)$.

The action \eqref{eq:action} defines our model for the superfluid droplet.
In the following subsections, we first clarify the role of the time-dependent Lagrange multiplier $\mu(t)$ enforcing the particle-number constraint, and then derive the equations of motion in the bulk together with the appropriate boundary condition at the moving interface.

\subsection{Noether current and particle-number constraint}
\label{sec:constraint}

To clarify the role of the Lagrange multiplier $\mu(t)$, we first examine the symmetry of the action \eqref{eq:action} and the associated Noether current.
Note that the action \eqref{eq:action} depends on the phonon field $\varphi$ only through its derivatives and is therefore invariant under the shift $\varphi \to \varphi + \alpha$ with a constant $\alpha$.
This invariance reflects the global $\U(1)$ symmetry that is spontaneously broken in the superfluid phase.
The associated Noether current $J^\mu$, where the spacetime index $\mu$ runs over $(t,r,\theta,\phi)$, is given by
\begin{equation}
 J^\mu := - \frac{\partial p}{\partial (\partial_\mu \varphi)}
 = 
 \begin{pmatrix}
   n (X) \\
   \frac{n (X)}{M} g^{ij} \partial_j \varphi
 \end{pmatrix},
 \label{eq:Noether-current}
\end{equation}
where we define the particle-number density as 
\begin{equation}
 n (X) := \left. \frac{\partial p (\mu)}{\partial \mu} \right|_{\mu = X}.
\end{equation}

We now turn to the role of the chemical potential.
As mentioned in the previous subsection, the chemical potential $\mu(t)$ serves as a time-dependent Lagrange multiplier that enforces the constraint on the total particle number.
This is seen by varying the action \eqref{eq:action} with respect to $\mu(t)$, which yields the constraint
\begin{equation}
 \int_{r<R} \diff^3 x \sqrt{g} n(X) = N = \const .
 \label{eq:constraint}
\end{equation}
Since the boundary location $r = R(t,\theta,\phi)$ is time dependent, the condition \eqref{eq:constraint} determines the time dependence of the chemical potential $\mu(t)$ in terms of the interface deformation $R(t,\theta,\phi)$.
We will demonstrate this relation explicitly in subsequent sections.

We emphasize that the introduction of the Lagrange multiplier, together with the phonon field $\varphi$, introduces a gauge redundancy under time-dependent shifts of the Nambu--Goldstone mode (see Appendix~\ref{sec:redundancy}). 
This redundancy allows one to fix the spatially uniform part of $\varphi$ with an appropriate gauge choice.
For example, one may impose the condition
\begin{equation}
 \int_{r<R} \diff^3 x\,\sqrt{g} \varphi (x) = 0.
 \label{eq:fixed-redundancy}
\end{equation}

\subsection{Equation of motion and boundary condition for the bulk superfluid}
\label{sec:eom}

Applying the variational principle to the effective action \eqref{eq:action}, we derive the equations of motion for the superfluid phonon field $\varphi$.
Special care is required in performing the variation, since the location of the interface, $R(t,\theta,\phi)$, depends explicitly on time $t$ and on the angular coordinates $(\theta,\phi)$.
As a result, variations of the bulk fields generate boundary contributions that must be treated carefully (see, e.g., Ref.~\cite{Watanabe:2014zza} for a related analysis in the case of an extended flat boundary).

First, since $X$ depends only on derivatives of the phonon field $\varphi$, the variation of the action with respect to $\varphi$ is given by
\begin{align}
 \delta S
 &= \int_{r<R} \diff t \diff^3 x \sqrt{g} 
 \frac{\partial p}{\partial (\partial_\mu \varphi) }\partial_\mu \delta \varphi
 \notag \\
 &= \int_{r<R} \diff t \diff^3 x \sqrt{g} 
 (-J^\mu) \partial_\mu \delta \varphi
 \label{eq:delta-S-0}
\end{align}
where we used the definition of the Noether current \eqref{eq:Noether-current}. 
Here and in the following, summation over repeated spacetime indices $\mu,\nu,\ldots=(t,r,\theta,\phi)$ is implied.

We now perform integration by parts in Eq.~\eqref{eq:delta-S-0}, taking into account that the integration domain is bounded by the time- and angle-dependent surface $r=R(t,\theta,\phi)$.
When integrating by parts with respect to $x^\mu=(t,r,\theta,\phi)$, derivatives acting on the integration region generate additional boundary contributions.
In particular, integration by parts with respect to the radial coordinate produces surface terms at the moving interface.
As a result, the variation of the action takes the form
\begin{align}
 \delta S
 =& \int_{r<R} \diff t \diff^3 x \sqrt{g}
 \delta \varphi \frac{1}{\sqrt{g}} \partial_\mu (\sqrt{g} J^\mu) 
 \notag 
 \\
 & 
 - \int \diff t \diff \Omega\,
 r^2
 \delta \varphi
 \bigl(J^r - J^{\mu} \partial_\mu R \bigr)
 \bigg|_{r=R(t,\theta,\phi)} .
 \label{eq:delta-S}
\end{align}

Requiring the variational principle $\delta S = 0$ yields the equations of motion in the bulk and the boundary condition at the interface:
\begin{align}
 0 
 &= \frac{1}{\sqrt{g}} \partial_\mu (\sqrt{g} J^\mu) ,
 \label{eq:conservation-law}
 \\
 0 &= J^r - J^{\mu}\partial_{\mu}R \Big|_{r=R} .
 \label{eq:boundary-condition}
\end{align}
Note that Eq.~\eqref{eq:conservation-law} represents the conservation law in the bulk because $\frac{1}{\sqrt{g}} \partial_\mu (\sqrt{g} J^\mu) = \partial_t J^0+\nabla_i J^i$ holds for the covariant derivative $\nabla_i$ associated with the metric in Eq.~\eqref{eq:metric}.
In contrast, the boundary condition~\eqref{eq:boundary-condition} admits a simple physical interpretation: it enforces particle-number conservation at the moving interface by properly accounting for the particle flux induced by the interface motion.
It is also worth emphasizing that substituting the explicit expression for the current $J^\mu$ into Eq.~\eqref{eq:conservation-law} generally leads to a nonlinear equation of motion.

\section{Fluctuations around a spherical background}
\label{sec:fluctuation}

The action \eqref{eq:action}, which already incorporates the particle-number constraint \eqref{eq:constraint}, provides a nonlinear description of superfluid phonons and surface oscillations.
At low energies, nonlinear terms necessarily involve additional derivatives and are therefore subleading in the derivative expansion.
We thus focus on small fluctuations around a static spherical background and formulate the corresponding quadratic theory, which forms the basis for the analysis of surface oscillations.
In Sec.~\ref{sec:background}, we first describe the spherical background configuration of our model.
In Sec.~\ref{sec:linearization}, we derive the linearized equations of motion and boundary conditions.
Finally, in Sec.~\ref{sec:quadratic-action}, we construct the effective quadratic action governing the phonon field $\varphi$ and the radial displacement $u$.

\subsection{Spherical background}
\label{sec:background}
We first specify the static spherical background, in which the fields take the configuration
\begin{equation}
 R = R_0, \quad \varphi = 0, \quad \mu = \mu_0.
\end{equation}
The corresponding background pressure and particle-number density are denoted as
\begin{equation}
 \pp := p(\mu_0),\quad  \bar{n} := n(\mu_0)=\frac{\partial p(\mu_0)}{\partial \mu_0}.
\end{equation}
With these expressions, Eq.~\eqref{eq:constraint} reduces to
\begin{equation}
 \bar{n} = \frac{N}{\Vcal (R_0)}
 \label{eq:static-number}
\end{equation}
with the volume $\Vcal (R_0)=\frac{4\pi}{3}R_0^3$ of the droplet.
Meanwhile, the variation of the action with respect to $R_0$ results in
\begin{equation}
  \pp\frac{\partial \Vcal(R_0)}{\partial R_0} = \sigma\frac{\partial \Acal[R_0]}{\partial R_0}
  \quad\Rightarrow\quad
  \pp=\frac{2\sigma}{R_0},
 \label{eq:Young-Laplace}
\end{equation}
which reproduces the Young--Laplace equation for a droplet in vacuum~\cite{Landau:Statistical,Landau:Fluid}.

Equations~\eqref{eq:static-number} and \eqref{eq:Young-Laplace} together determine the background chemical potential $\mu_0$ and the radius $R_0$ of the droplet for a given total particle number $N$.
The spherical configuration specified above serves as the reference background around which we study small fluctuations of the interface and the superfluid phonon field in the following subsection.

\subsection{Linearized dynamics}
\label{sec:linearization}

We now consider the linearized dynamics around the static spherical background specified in the previous subsection.
In this analysis, the chemical potential $\mu(t)$ is kept as a time-dependent Lagrange multiplier that enforces the fixed particle-number constraint.
Accordingly, we write
\begin{equation}
 \mu(t) = \mu_0 + \delta \mu(t),
\end{equation}
where $\delta \mu(t)$ is determined by the particle-number constraint.
In the following, we treat $\delta \mu(t)$, $\varphi(x)$, and $u(t,\theta,\phi)$ as small fluctuations.

The stationary conditions~\eqref{eq:static-number} and \eqref{eq:Young-Laplace} ensure that, upon expanding the action around the background configuration, no term linear in $\delta \mu$ alone appears.
Mixed terms involving $\delta \mu(t)$ and the fluctuation fields, however, remain and lead to the particle-number constraint at the level of small fluctuations.
To find the particle-number constraint for small fluctuations, we linearize Eq.~\eqref{eq:constraint}.

In the present case, $n(X)$ is expanded around the ground sate as
\begin{align}
 n(X) \simeq \bar{n} + \chi (\delta \mu - \partial_t \varphi),
\end{align}
where $\chi$ is the compressibility evaluated on the background configuration
\begin{align}
 \chi
 :=\frac{\partial \bar{n}}{\partial \mu_0}
 =\frac{\partial^2 \bar{p}}{\partial \mu_{0}^2}.
\end{align}
At the same time, the integration domain also provides corrections as
\begin{align}
\int_{r<R} \diff^3 x \sqrt{g} 
 \simeq \int_{r<R_0} \diff^3 x \sqrt{g} + \int \diff \Omega\, R_0^2 u. \label{eq:correction-integration-domain}
\end{align}
As a result, Eq.~\eqref{eq:constraint} for small fluctuations yields
\begin{equation}
 0 \simeq 
 \chi\int_{r< R_0} \diff^3 x\,\sqrt{g} (\delta \mu - \partial_t \varphi)
 + \bar{n} R_0^2\int \diff \Omega\,  u,
\end{equation}
where we used Eq.~\eqref{eq:static-number}.
Using the gauge-fixing condition~\eqref{eq:fixed-redundancy} at linear order, the spatially uniform part of $\varphi$ is removed, so that the term proportional to $\partial_t \varphi$ vanishes, resulting in
\begin{equation}
 \delta \mu (t)
 = - \frac{\bar{n} R_0^2}{\chi \Vcal (R_0)}
 \int \diff \Omega\, u(t,\theta,\phi).
 \label{eq:delta-mu1}
\end{equation}

We next derive the linearized equations of motion and the corresponding boundary conditions for the phonon field $\varphi$.
To this end, we linearize the Noether current as
\begin{equation}
 J^\mu
 =
 \begin{pmatrix}
   \bar{n} - \chi (\partial_t \varphi - \delta \mu)\\
   \frac{\bar{n}}{M} g^{ij} \partial_j \varphi
 \end{pmatrix}
 + O \big( (\partial \varphi)^2 \big) .
 \label{eq:linearized-CR}
\end{equation}
Substituting this linearized current into the continuity equation \eqref{eq:conservation-law} and into the boundary condition~\eqref{eq:boundary-condition}, we obtain the following set of linearized equations
\begin{align}
 - \chi \partial_t^2 \varphi
 + \frac{\bar{n}}{M} \bnab^2 \varphi
 &= - \chi \partial_t \delta \mu, 
  \label{eq:linearized-eom-0} \\
 \partial_r \varphi \big|_{r=R_0} 
 &= M \partial_t u,
 \label{eq:linearized-boundary-cond}
\end{align}
with the Laplacian in spherical coordinates, $\bnab^2 = \frac{1}{\sqrt{g}} \partial_i (\sqrt{g} g^{ij} \partial_j )$, associated with the metric~\eqref{eq:metric}.
Equation~\eqref{eq:linearized-boundary-cond} represents the linearized kinematic boundary condition, stating that the normal velocity of the interface matches the radial component of the superfluid velocity at the boundary.

The equation of motion for $\varphi$ apparently contains a source term proportional to $\partial_t \delta\mu$.
However, this term is spatially uniform and therefore couples only to the position-independent component of $\varphi$.
Since the gauge-fixing condition~\eqref{eq:fixed-redundancy} removes this mode, the source term does not affect the physical phonon excitations.
In fact, integrating Eq.~\eqref{eq:linearized-eom-0} over the entire superfluid droplet and using the gauge-fixing condition together with the boundary condition~\eqref{eq:linearized-boundary-cond}, one finds that it merely reproduces Eq.~\eqref{eq:delta-mu1} and does not impose any additional constraint.
Consequently, the equation of motion for the physical phonon excitations reduces to a homogeneous wave equation:
\begin{equation}
 - \chi \partial_t^2 \varphi 
 + \frac{\bar{n}}{M} \bnab^2 \varphi = 0,
 \label{eq:linearized-eom} 
\end{equation}
which takes the familiar form of a wave equation with the sound velocity $c_s := \sqrt{\bar{n}/(M \chi)}$.

\subsection{Quadratic action}
\label{sec:quadratic-action}

We next expand the action~\eqref{eq:action} around the static spherical background in powers of the fluctuations of the superfluid phonon field $\varphi$ and the radial displacement $u$ defined in Eq.~\eqref{eq:radius-R}.
Keeping terms up to quadratic order captures the leading low-energy dynamics of small fluctuations.

We first expand the bulk Lagrangian density $p(X)$ around the background chemical potential $\mu_0$, yielding
\begin{align}
 p (X) 
 &\simeq
 \pp
 + \bar{n}
 \left[
  \delta\mu - \partial_t \varphi - \frac{(\bnab \varphi)^2}{2M}
 \right]
 \notag \\
 &\quad + \frac{\chi}{2}
 \left( \delta\mu - \partial_t \varphi \right)^2 ,
 \label{eq:pn-expansion}
\end{align}
with $\delta \mu$ given in Eq.~\eqref{eq:delta-mu1}.
Substituting Eq.~\eqref{eq:pn-expansion} into the action~\eqref{eq:action} and combining it with the correction term in Eq.~\eqref{eq:correction-integration-domain} arising from the expansion of the integration domain, we retain terms up to quadratic order in the fluctuations and perform integrations by parts to separate bulk and boundary contributions.
Keeping the nonvanishing contributions, including boundary terms generated by integrations by parts, we obtain
\begin{widetext}
\begin{align}
 S
 =&\, \pp \int \diff t\, \Vcal [R]
  - \sigma \int \diff t\, \Acal [R]
  + \int \diff t  
  \left[
   \frac{\chi \Vcal (R_0)}{2} (\delta \mu)^2
   + \bar{n} R_0^2 \delta \mu \int \diff \Omega u
  \right]
  \notag  \\
  &- \int \diff t \diff \Omega\, 
  R_0^2
  \left[
   \varphi 
   \left( 
    \frac{\bar{n}}{2M} \partial_r \varphi
   - \bar{n} \partial_t u
   \right)
  \right]
  \bigg|_{r=R_0}
+ \frac{1}{2} \int_{r<R_0} \diff t \diff^3 x 
  \sqrt{g}
  \varphi
  \left[
    - \chi \partial_t^2 \varphi
    + \frac{\bar{n}}{M} \bnab^2 \varphi
  \right],
 \label{eq:quadratic-action-0}
\end{align}
where we introduced the droplet volume
\begin{equation}
 \Vcal [R] := \frac{1}{3} \int \diff \Omega\, R^3
\end{equation}
Here, the linear terms in the fluctuations vanish due to the equilibrium conditions, and the mixed term proportional to $\delta \mu \partial_t \varphi$ vanishes upon imposing the gauge-fixing condition~\eqref{eq:fixed-redundancy}. 
Varying the action with respect to $\delta \mu$ reproduces Eq.~\eqref{eq:delta-mu1}, which encodes the particle-number constraint.

We can further simplify the geometric part of Eq.~\eqref{eq:quadratic-action-0} by expanding around $R(t,\theta,\phi)=R_0+u(t,\theta,\phi)$, using the equilibrium condition~\eqref{eq:Young-Laplace}, and eliminating $\delta\mu$ with Eq.~\eqref{eq:delta-mu1}.
Carrying out this procedure, we obtain the quadratic effective action
\begin{equation}
 S^{(2)} 
 = S_{\mathrm{geom}} + S_{\mathrm{bdry}} + S_{\mathrm{bulk}} ,
 \label{eq:quadratic-action}
\end{equation}
where each term corresponds to the respective line of Eq.~\eqref{eq:quadratic-action-0}:
\begin{align}
 S_{\mathrm{geom}}
 &:= - \frac{\sigma}{2} \int \diff t \diff \Omega
  u \left( \hat{\bm{L}}^2 - 2\right) u
  - \frac{3 \bar{n}^2 R_0}{8 \pi \chi} 
  \int \diff t \left( \int \diff \Omega\, u \right)^2 ,
 \label{eq:geom-action-u} 
 \\
 S_{\mathrm{bdry}}
 &:= - \int \diff t \diff \Omega 
  R_0^2
  \left[
   \varphi
   \left( 
    \frac{\bar{n}}{2M} \partial_r \varphi
   - \bar{n} \partial_t u
   \right)
  \right]
  \bigg|_{r=R_0},
 \label{eq:bdry-action} 
 \\
 S_{\mathrm{bulk}}
 &:= \frac{1}{2} \int_{r<R_0} \diff t \diff^3 x 
  \sqrt{g}
  \varphi
  \left[
    - \chi \partial_t^2 \varphi
    + \frac{\bar{n}}{M} \bnab^2 \varphi
  \right].
 \label{eq:bulk-action} 
\end{align}
\end{widetext}
Here, we dropped an irrelevant constant term and introduced the orbital angular momentum operator $\hat{\bm L}^2$ on the unit sphere:
\begin{align}
 \hat{\bm{L}}^2
 = -\frac{1}{\sin\theta}
 \frac{\partial}{\partial \theta}
 \left(
 \sin\theta
 \frac{\partial}{\partial \theta}
 \right)
 - \frac{1}{\sin^2\theta}
 \frac{\partial^2}{\partial\phi^2}.
\end{align}

The second term in $S_{\mathrm{geom}}$ arises from integrating out the time-dependent Lagrange multiplier $\delta\mu(t)$ associated with the global particle-number constraint.
This term is nonlocal on the sphere and affects only volume-changing deformations, namely the breathing mode.
It therefore provides a restoring force for the breathing mode and effectively generates a mass term for this mode, as will be discussed in the next section.

In contrast, the bulk term $S_{\mathrm{bulk}}$ is proportional to the linearized equation of motion for the phonon field and therefore vanishes when evaluated on shell.
The variation of $S_{\mathrm{bulk}}$ reproduces the homogeneous linearized equation of motion~\eqref{eq:linearized-eom}, consistent with the gauge-fixing condition imposed in deriving the quadratic action.
Consequently, the geometric and boundary contributions, $S_{\mathrm{geom}}$ and $S_{\mathrm{bdry}}$, fully determine the effective theory for surface oscillations.

\section{Surface dynamics of a self-bound superfluid droplet}
\label{sec:vacuum}

Based on the formulation developed in the previous section, we now investigate surface oscillations of a self-bound superfluid droplet.
In Sec.~\ref{sec:surfaceEFT-in-vacuum}, we derive the effective action governing small-amplitude surface oscillations.
In Sec.~\ref{sec:spectrum-in-vacuum}, we analyze the normal-mode frequencies both analytically and numerically.
In Sec.~\ref{sec:quantization-in-vacuum}, we apply canonical quantization to the low-energy surface oscillations and discuss the resulting quanta, known as ripplons.

\subsection{Effective action for surface oscillations}
\label{sec:surfaceEFT-in-vacuum}

The effective theory for surface oscillations is obtained by solving the linearized bulk equation~\eqref{eq:linearized-eom} subject to the boundary condition~\eqref{eq:linearized-boundary-cond}, and then substituting the on-shell solution into the quadratic action~\eqref{eq:quadratic-action}.

\begin{figure*}[htb]
  \centering
  \includegraphics[trim={1cm 6cm 1cm 0cm}, clip, width=0.8\linewidth]{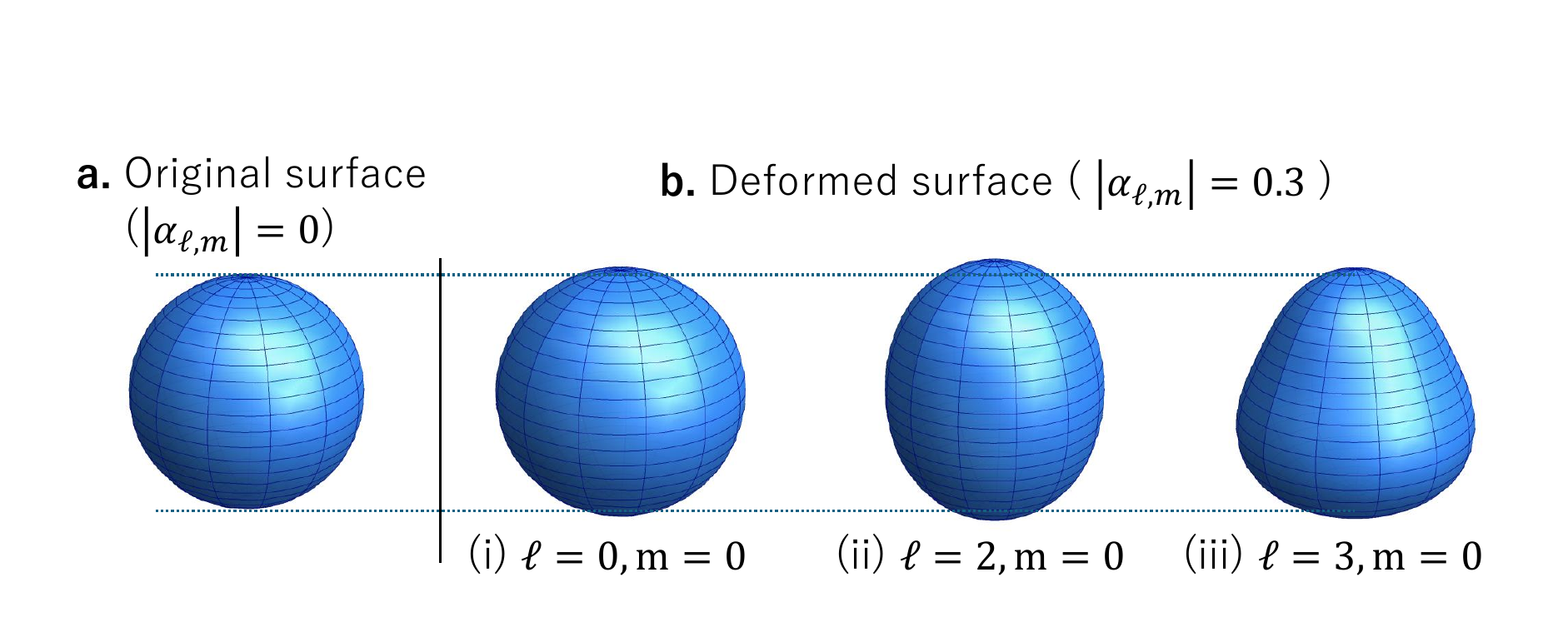}
  \caption{
  Schematic illustration of the surface deformation modes.
  (a) The original surface ($|\alpha_{\ell,m}|=0$).
  (b) Deformed surface with $|\alpha_{\ell,m}|=0.3$, illustrating
  (i) the breathing mode ($\ell=0, m=0$),
  (ii) the quadrupole mode ($\ell=2, m=0$), and
  (iii) the octupole mode ($\ell=3, m=0$).
  }
    \label{fig:surface-deformation}
\end{figure*}

We begin by solving the linearized equations~\eqref{eq:linearized-eom} and \eqref{eq:linearized-boundary-cond}.
Since the background configuration is spherically symmetric, it is natural to decompose the fields into spherical harmonics labeled by the angular momentum quantum numbers $(\ell,m)$.
Accordingly, we expand the phonon field and the radial displacement as
\begin{subequations}\label{eq:expansion-field-vacuum}
\begin{align}
 \varphi (t,\bx)
 &=
 \sum_{\ell = 0}^\infty \sum_{m=-\ell}^\ell
 \varphi_{\ell m}(t,r)\,
 Y_{\ell m} (\theta,\phi),
 \\
 u (t,\theta,\phi)
 &=
 R_0
 \sum_{\ell = 0}^\infty \sum_{m=-\ell}^\ell
 \alpha_{\ell m}(t)\,
 Y_{\ell m} (\theta,\phi).
\end{align}
\end{subequations}
The reality condition for $u(t,\theta,\phi)$ requires $\alpha_{\ell m}^\ast(t)=(-1)^m \alpha_{\ell,-m}(t)$.
Using Eq.~\eqref{eq:radius-R}, we obtain
\begin{equation}
 R (t,\theta,\phi)
 = R_0 
 \left[ 
  1 + \sum_{\ell = 0}^\infty \sum_{m=-\ell}^\ell
 \alpha_{\ell m}(t)
 Y_{\ell m} (\theta,\phi)
 \right],
\end{equation}
showing that $\alpha_{\ell m}(t)$ represents the dimensionless amplitudes of the surface deformation modes (see Fig.~\ref{fig:surface-deformation}).

Our goal in this subsection is to derive the effective action in terms of $\alpha_{\ell m}(t)$.
Since the background is invariant under time translations, it is convenient to Fourier transform in time.
We therefore express the above mode expansions in Fourier space as
\begin{subequations}
\begin{align}
 \varphi (t,\bx)
 &=
 \int \frac{\diff \omega}{2\pi}
 \rme^{-\rmi \omega t}
 \sum_{\ell = 0}^\infty \sum_{m=-\ell}^\ell
 \tilvarphi_{\ell m}(\omega,r)
 Y_{\ell m}(\theta,\phi),
 \\
 u(t,\theta,\phi)
 &=
 R_0
 \int \frac{\diff \omega}{2\pi}
 \rme^{-\rmi \omega t}
 \sum_{\ell = 0}^\infty \sum_{m=-\ell}^\ell
 \tilalpha_{\ell m}(\omega)
 Y_{\ell m}(\theta,\phi).
\end{align}
\end{subequations}
The reality condition for $u(t,\theta,\phi)$ implies $\tilalpha_{\ell m}^\ast(\omega) = (-1)^m \tilalpha_{\ell,-m}(-\omega)$.

With this expansion, we first simplify the geometric contribution $S_{\mathrm{geom}}$ of the action.
Using the angular integral of the spherical harmonics, we obtain
\begin{equation}
 \int \diff \Omega\, u 
 = \sqrt{4\pi} R_0 \int \frac{\diff \omega}{2\pi} \rme^{-\rmi \omega t}
 \tilalpha_{00} (\omega).
\end{equation}
Further using $\hat{\bm{L}}^2 Y_{\ell m} = \ell (\ell+1) Y_{\ell m}$ and the orthonormality of the  spherical harmonics, we rewrite $S_{\mathrm{geom}}$ as 
\begin{widetext}
\begin{equation}
 S_{\mathrm{geom}}
 = - \frac{1}{2} 
 \int \frac{\diff \omega}{2\pi}
 \sum_{\ell = 0}^\infty \sum_{m=-\ell}^{\ell} 
 \left[ 
  \sigma R_0^2 (\ell-1) (\ell+2) 
  + \frac{3 \bar{n}^2 R_0^3}{\chi} \delta_{\ell 0} \delta_{m0}
 \right]|\tilalpha_{\ell m} (\omega)|^2.
 \label{eq:geometry-action-1}
\end{equation}
where $\delta_{ij}$ denotes the Kronecker delta.
\end{widetext}

There are two salient features of the geometric contribution.
First, the term originating from the particle-number constraint contributes only to the $\ell=0$, $m=0$ mode, namely the breathing mode. 
This reflects the fact that fixing the total particle number couples exclusively to volume-changing deformations.
As a result, this contribution plays an essential role in stabilizing the droplet. 
If instead one were to fix the chemical potential and omit the particle-number constraint, the geometric part of the action would contain only the surface-tension term, which by itself favors a reduction of the droplet radius and thus signals an instability of the self-bound configuration. 
As we will see later, however, the breathing mode can still be stabilized, even in the fixed-chemical-potential setup, provided that the surface tension does not exceed a critical value. 
This stabilization arises because the boundary action $S_{\mathrm{bdry}}$ induces an additional positive restoring force.

Second, we note that the $\ell=1$ modes do not contribute to $S_{\mathrm{geom}}$. 
These modes correspond to rigid translations of the droplet and therefore do not generate any restoring force. 
This property is a direct consequence of the self-bound nature of the system and contrasts with droplets confined by an external trapping potential, where the translational modes acquire a finite frequency.

We next solve for the bulk phonon field induced by the surface motion.
With the mode expansion in Eq.~\eqref{eq:expansion-field-vacuum}, the linearized bulk equation~\eqref{eq:linearized-eom} and the boundary condition~\eqref{eq:linearized-boundary-cond} are reduced to
\begin{align}
 &0 = 
 \frac{\diff^2 \tilvarphi_{\ell m}}{\diff r^2}
 + \frac{2}{r} \frac{\diff \tilvarphi_{\ell m}}{\diff r} 
 + \left( \frac{\omega^2}{c_s^2} - \frac{\ell (\ell+1)}{r^2} \right) \tilvarphi_{\ell m},
 \label{eq:Bessel-ODE}
 \\
 &\left. 
  \frac{\partial \tilvarphi_{\ell m} (\omega,r) }{\partial r}
 \right|_{r=R_0}
 = - \rmi \omega M R_0 \tilalpha_{\ell m} (\omega).
 \label{eq:boundary-condition-Fourier}
\end{align}
The general solution of Eq.~\eqref{eq:Bessel-ODE} is a linear combination of the spherical Bessel functions of the first and second kind,
\begin{equation}
 \tilvarphi_{\ell m} (\omega,r) 
 = A_{\ell m} j_{\ell} \left(\omega r/c_s\right)
  + B_{\ell m} y_{\ell} (\omega r/c_s) .
\end{equation}
Imposing regularity at the origin $r=0$ requires $B_{\ell m}=0$.  
The boundary condition~\eqref{eq:boundary-condition-Fourier} then determines the remaining coefficient as
\begin{equation}
 A_{\ell m}
 = - \rmi f_\ell(\omega) \tilalpha_{\ell m}(\omega),
 \quad
 f_\ell(\omega)
 := \frac{c_s M R_0}{j'_{\ell}(\frac{\omega R_0}{c_s})},
\end{equation}
where $j'_\ell(x) := \diff j_\ell(x)/\diff x$.
We also note that $f_{\ell}(-\omega)=(-1)^{\ell+1}f_{\ell}(\omega)$, which follows from the parity property $j_{\ell}(-z)=(-1)^{\ell}j_{\ell}(z)$.
As a result, the bulk phonon field can be expressed in terms of the surface deformation amplitudes as
\begin{align}
 \varphi (t,\bx)
 = \int \frac{\diff \omega}{2\pi}\,
   \rme^{-\rmi \omega t}
 \sum_{\ell=0}^\infty 
 &\sum_{m=-\ell}^{\ell}
 \bigl[-\rmi f_\ell(\omega)\, \tilalpha_{\ell m}(\omega)\bigr]
 \notag \\
 &\times j_{\ell}(\omega r/c_s)
  Y_{\ell m}(\theta,\phi).
\end{align}

We then substitute this solution back into the quadratic action~\eqref{eq:quadratic-action}.
Since the bulk contribution~\eqref{eq:bulk-action} is proportional to the equations of motion and hence vanishes on shell, we only need to evaluate the boundary contribution~\eqref{eq:bdry-action}.
After an explicit calculation, using the relation $c_s^2 = \bar{n}/(M \chi)$, we obtain
\begin{equation}
 S_{\mathrm{bdry}}
 = \frac{1}{2} \int \frac{\diff \omega}{2\pi} \sum_{\ell=0}^\infty \sum_{m=-\ell}^{\ell}
  \frac{\bar{n}^2 R_0^3}{\chi}
 \frac{\frac{\omega R_0}{c_s} j_{\ell} (\frac{\omega R_0}{c_s}) }{j_{\ell}' (\frac{\omega R_0}{c_s})}
 |\tilalpha_{\ell m} (\omega)|^2 .
 \label{eq:boundary-action-1}
\end{equation}
Combining the geometric contribution with the boundary term~\eqref{eq:boundary-action-1}, we arrive at the effective action for surface oscillations, $S_{\eff} = S_{\mathrm{eff}} [\alpha_{\ell m}]$, given by
\begin{widetext}
\begin{equation}
 S_{\eff} 
 = \frac{1}{2} \int \frac{\diff \omega}{2\pi} \sum_{\ell=0}^\infty \sum_{m=-\ell}^\ell
 \left[
 \frac{\bar{n}^2 R_0^3}{\chi}
 \frac{\frac{\omega R_0}{c_s} j_{\ell} (\frac{\omega R_0}{c_s}) }{j_{\ell}' (\frac{\omega R_0}{c_s})}
  - 
  \left( 
   \sigma R_0^2 (\ell-1) (\ell+2) 
   + \frac{3 \bar{n}^2 R_0^3}{\chi} \delta_{\ell 0} \delta_{m0}
  \right)
 \right] |\tilalpha_{\ell m} (\omega)|^2 .
 \label{eq:action-surface-1}
\end{equation}
This effective action governs the dynamics of each surface oscillation mode $\alpha_{\ell m}$.
\end{widetext}

\subsection{Normal-mode spectrum}
\label{sec:spectrum-in-vacuum}

Since the effective action~\eqref{eq:action-surface-1} is diagonal in the angular momentum indices $\ell$ and $m$, the normal-mode analysis can be carried out independently for each $(\ell,m)$ mode.

The on-shell condition for surface oscillations is then obtained from the kernel appearing in the action \eqref{eq:action-surface-1} as
\begin{align}
 0 =& 
 \frac{\bar{n}^2 R_0^3}{\chi}
 \frac{\frac{\omega_{\ell} R_0}{c_s} j_{\ell} (\frac{\omega_{\ell} R_0}{c_s}) }{j_{\ell}' (\frac{\omega_{\ell} R_0}{c_s})}
 \notag \\
 & - 
  \left( 
   \sigma R_0^2 (\ell-1) (\ell+2) 
   + \frac{3 \bar{n}^2 R_0^3}{\chi} \delta_{\ell 0} \delta_{m0}
  \right).
\end{align}
As anticipated, the above equation does not depend on $m$ due to the spherical symmetry of the background, and the normal mode frequency depends only on $\ell$, which we denote by $\omega_{\ell}$.
Separating the breathing mode ($\ell=0$) from the higher multipole modes ($\ell \ge 1$) and introducing the dimensionless normal-mode frequency 
\begin{equation}
 z_{\ell} := 
 \frac{\omega_{\ell} R_0}{c_s}
 ,
\end{equation}
the condition reduces to
\begin{subequations}\label{eq:frequency-condition}
\begin{align}
 \frac{z_0  j_0 (z_0) }{j_0' (z_0)}
 &= 3 - 2 \xi \quad
 \\
 \frac{z_{\ell} j_{\ell} (z_{\ell}) }{j_{\ell}' (z_{\ell})}
 &= (\ell-1) (\ell+2) \xi \quad (\ell \geq 1).
 \end{align}
\end{subequations}
Here, we have introduced the dimensionless parameter 
\begin{equation}
 \xi := \frac{\chi \sigma}{\bar{n}^2 R_0} 
 = \frac{\chi \pp}{2 \bar{n}^2}\label{eq:dimless-parameter}
 ,
\end{equation}
where we used the Young--Laplace equation \eqref{eq:Young-Laplace} to obtain the rightmost expression.

This dimensionless parameter $\xi$ characterizes the relative importance of surface tension and bulk compressibility in determining the oscillation spectrum. 
In the thin-wall limit, the normal-mode frequencies depend on the microscopic details only through this parameter.
To specify the value of $\xi$, we need information about the finite droplet, such as the surface tension $\sigma$ and the ground-state radius $R_0$, or equivalently, the pressure in the ground state $\pp$.
Using $c_s^2=\bar n/(M\chi)$, we can rewrite $\xi$ as
\begin{align}
 \xi =
  \frac{\chi\,\sigma}{\bar n^{\,2}R_0}
 =\frac{1}{R_0}\,\frac{\sigma}{M\bar n c_s^2}
 = \frac{l_\sigma}{R_0},
 \quad
 l_\sigma:=\frac{\sigma}{M\bar n c_s^2},
 \label{eq:xi-another-expression}
\end{align}
where we introduced the length scale $l_\sigma$, which plays the role of a capillary length.
This expression makes it clear that information about the finite droplet is required to determine all coefficients entering the spectrum discussed below.

Identification of the parameters in our model requires additional input either from microscopic theoretical calculations or from experiments.
For example, the surface tension can in principle be obtained microscopically by solving for the nonuniform interfacial profile under boundary conditions appropriate for a droplet and evaluating the excess energy per unit area associated with the interface~\cite{Ancilotto:2018}.
Other quantities, such as $\chi$ and $\pn$, are determined from their values in the thermodynamic limit, as we will show in Sec.~\ref{sec:application-cold-atom} in a concrete example.
In experiments, $\xi$ could in principle be extracted from the eigenfrequency of a surface mode, provided that the thin-wall description is reasonably accurate.

\paragraph{Numerical results.}
We numerically solve the frequency condition~\eqref{eq:frequency-condition} to obtain the dispersion relation for the surface modes.

\begin{figure*}
\centering
\includegraphics[trim={3.5cm 1.8cm 5.5cm 0cm}, clip, width=1\linewidth]{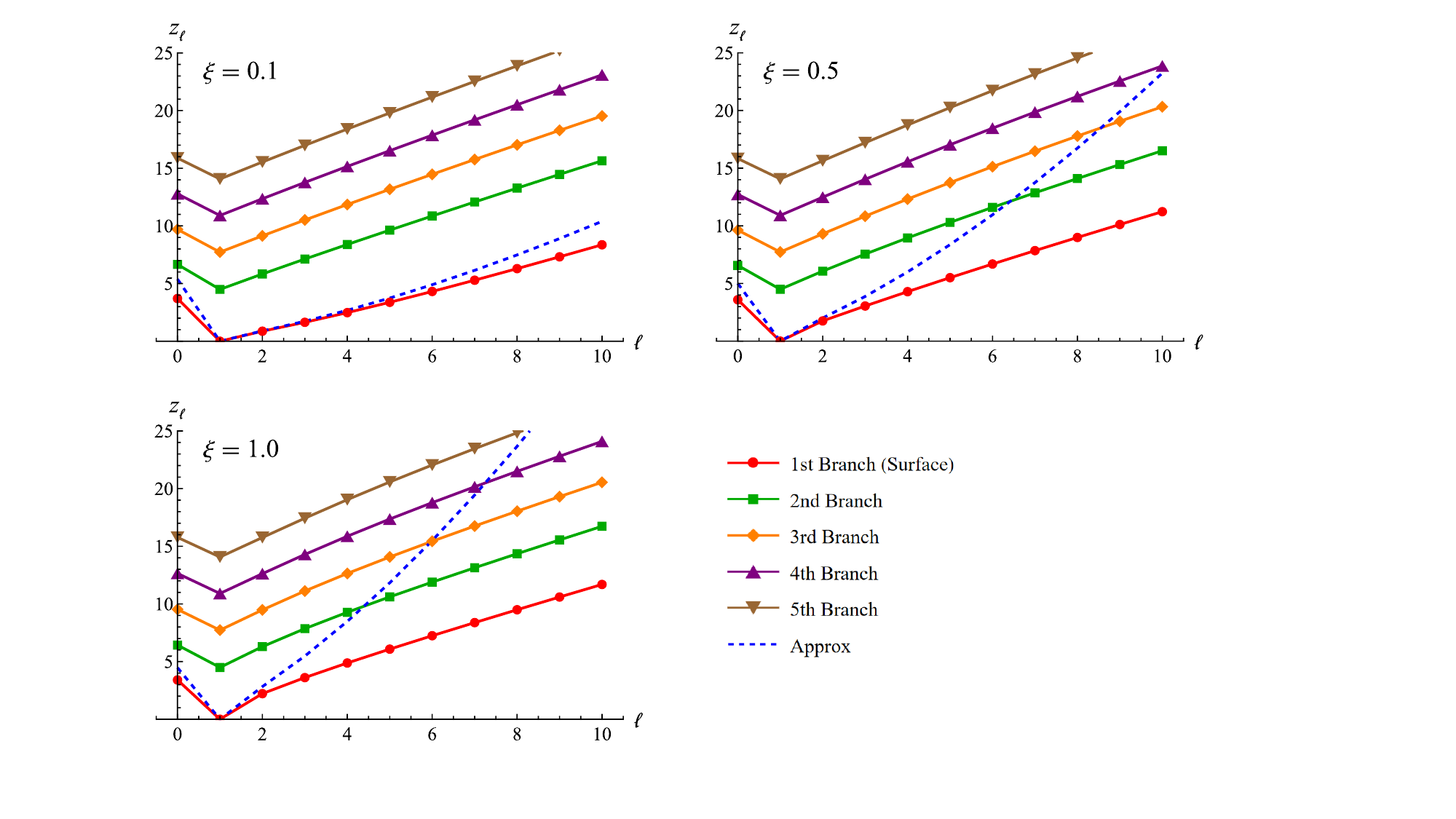}
    \caption{
    Numerical solutions to Eq.~\eqref{eq:frequency-condition} of the dimensionless frequencies $z_\ell$ as functions of $\ell$ for $\xi=0.1$ (top-left), $\xi=0.5$ (top-right) and $\xi=1.0$ (bottom-left). 
    Different markers correspond to distinct normal-mode branches, and the dashed line indicates the low-frequency approximation given by Eq.~\eqref{eq:low-energy-freq}.}
    \label{fig:plot-zl-l}
\end{figure*}

\begin{figure*}
    \centering
    \includegraphics[trim={4cm 3.5cm 5cm 0.5cm}, clip, width=1.0\linewidth]{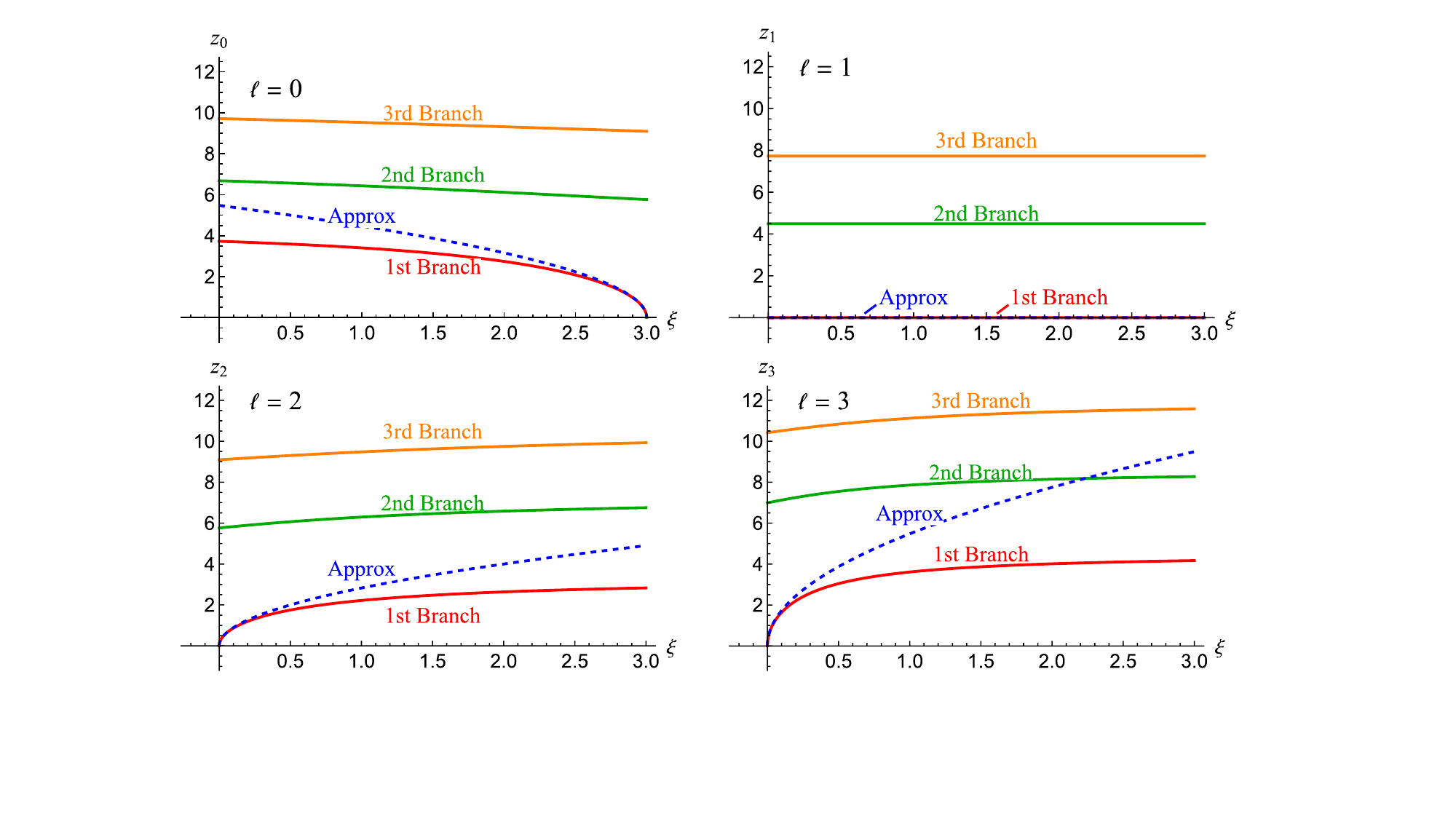}
    \caption{
    Numerical solutions of the dimensionless frequencies $z_\ell$ as functions of $\xi$ for $\ell=0$ (top-left) , $\ell=1$ (top-right) , $\ell=2$ (bottom-left) and $\ell=3$ (bottom-right).
    Different curves correspond to distinct normal-mode branches, and the dashed curve indicates the low-frequency approximation given by Eq.~\eqref{eq:low-energy-freq}].
    For $\ell=0$, the lowest branch vanishes at $\xi=3$.
    In the upper-right figure for $\ell=1$, both the first branched solution and the approximate solution are always zero and lie exactly on the same line.}
    \label{fig:plot-zl-xi}
\end{figure*}

Figure~\ref{fig:plot-zl-l} demonstrates numerical solutions of Eq.~\eqref{eq:frequency-condition} for the dimensionless frequencies $z_{\ell}$ at fixed $\ell$.
This determines the normal-mode eigenfrequency for each angular momentum.
The solutions for each value of $\xi$ are also shown in Fig.~\ref{fig:plot-zl-xi} as solid lines, while the approximate solutions discussed below are indicated by dashed lines.

We observe that the approximate solutions show good agreement with the numerical results only in the regime of small $z_\ell$.
Moreover, the full numerical solution exhibits multiple branches.
The first branch (the lowest-frequency mode) describes surface waves that do not possess nodes in the bulk.
In contrast, the higher-frequency branches correspond to standing waves inside the droplet, characterized by nodes in the bulk fluid.

\paragraph{Low-frequency approximation.}
In the low-frequency regime, $z_{\ell} = \omega_{\ell} R_0 / c_s \ll 1$, analytic expressions for the normal-mode frequencies can be derived.

In this limit, the spherical Bessel functions can be expanded for small arguments as
\begin{subequations}\label{eq:Bessel-expansion}
 \begin{align}
 \frac{z j_0 (z)}{j_0' (z)} 
 &= -3 + \frac{z^2}{5} + O(z^4)
 \\
 \frac{z j_{\ell} (z)}{j_{\ell}' (z)} 
 &= \frac{z^2}{\ell} + O(z^4) \quad (\ell \geq 1)
 \end{align}   
\end{subequations}
allowing us to expand the frequency-dependent kernel of the effective action in powers of $\omega$ as
\begin{equation}
 \begin{split}
 S_{\eff} 
 &\simeq \int \frac{\diff \omega}{2\pi}
 \sum_{\ell=0}^\infty \sum_{m=-\ell}^\ell
 \left[
 \frac{1}{2} B_{\ell} \omega^2 
 - \frac{1}{2} C_{\ell}
 \right] |\tilalpha_{\ell m} (\omega)|^2 ,
 \label{eq:surface-action-low-energy-frequency}
 \end{split}
\end{equation}
where we kept the terms up to $O(\omega^2)$ and introduced the mass parameter $B_{\ell}$ and elastic parameter $C_{\ell}$ as 
\begin{subequations}\label{eq:low-energy-parameters}
\begin{align}
 B_{\ell}
 &=
 \begin{cases}
  \dfrac{M \bar{n} R_0^5}{5} 
  \quad (\ell= 0),
  \vspace{5pt} \\
  \dfrac{M \bar{n} R_0^5}{\ell} 
  \quad (\ell\geq 1),
 \label{eq:mass-parameter-1} 
 \end{cases}
 \vspace{5pt} \\
 C_{\ell}
 &=
 \begin{cases}
 2 \left(
 \dfrac{3 \bar{n}^2 R_0^3}{\chi} 
 - \sigma R_0^2 
 \right)
 \quad (\ell =0),
 \vspace{5pt} \\
 \sigma R_0^2 (\ell-1) (\ell+2) 
 \quad (\ell\geq 1) .
 \end{cases} 
 \label{eq:elastic-parameter-1}
\end{align}    
\end{subequations}

Two remarks are in order.
First, the mass and elastic parameters~\eqref{eq:mass-parameter-1} and~\eqref{eq:elastic-parameter-1} for $\ell \geq 1$ coincide with those obtained in the classical incompressible liquid-drop model, well known from studies of surface oscillations in nuclear physics.
In contrast, the mass and elastic parameters for the breathing mode ($\ell=0$) are specific to the present theory, as they arise from the particle-number constraint and the bulk compressibility.

Second, Eq.~\eqref{eq:Bessel-expansion} implies that the low-frequency expansion of the kernel generates an additional restoring force for the breathing mode ($\ell=0$).
Because this dynamical contribution arises from the bulk phonon dynamics encoded in the boundary action, the breathing mode can remain stable even in a fixed-chemical-potential setup.
The nonzero breathing-mode frequency therefore reflects dynamical bulk responses of the superfluid, rather than being determined solely by geometric surface effects.

In the time domain, the low-energy effective action~\eqref{eq:surface-action-low-energy-frequency} can be written as
\begin{equation}
 S_{\eff} 
 \simeq \int \diff t 
 \sum_{\ell=0}^\infty \sum_{m=-\ell}^\ell
 \left[
  \frac{1}{2} B_{\ell} |\dot{\alpha}_{\ell m} (t)|^2 
  - \frac{1}{2} C_{\ell} |\alpha_{\ell m} (t)|^2
 \right] ,
 \label{eq:surface-action-low-energy-time}
\end{equation}
demonstrating that the low-energy dynamics reduces to an infinite set of decoupled harmonic oscillators.  
The corresponding eigenfrequencies are thus given by
\begin{equation}
 \omega_{\ell} := \sqrt{\frac{C_{\ell}}{B_{\ell}}} = 
 \begin{cases}
 \dfrac{c_s}{R_0} \sqrt{ 10 \left(3 - \xi \right)}
  ~~ (\ell =0),
  \vspace{5pt} \\
  \dfrac{c_s}{R_0}
  \sqrt{ \xi
  \ell (\ell-1) (\ell+2)}
  ~~ (\ell \geq 1).
 \end{cases}
 \label{eq:low-energy-freq}
\end{equation}
We also note that by substituting $c_s=\sqrt{\bar{n}/(M\chi)}$ and the definition of $\xi$, i.e., Eq.~\eqref{eq:dimless-parameter}, into the eigenfrequency $\omega_{\ell}$ for $\ell\geq 1$, one can express it as
\begin{align}
\omega_{\ell}=\sqrt{\frac{\sigma}{M \bar{n}R_{0}^3}\ell (\ell-1) (\ell+2)}
\quad 
(\ell \geq 1),
\end{align}
which coincides with the well-known result first derived by Lord Rayleigh for capillary oscillations of an incompressible liquid drop~\cite{Rayleigh1879}.
The validity of the low-energy expansion depends on the angular-momentum sector through the small parameter $z_\ell = \omega_\ell R_0 / c_s$.
In particular, the breathing mode becomes soft as $\xi$ approaches the critical value $\xi_{\mathrm{cr}} := 3$, whereas the higher multipole modes are controlled by the smallness of $\xi$.

In Figs.~\ref{fig:plot-zl-l} and~\ref{fig:plot-zl-xi}, we compare the analytic low-energy results~\eqref{eq:low-energy-freq} (shown as dashed lines) with the numerical solutions of Eq.~\eqref{eq:frequency-condition} (solid lines).
As expected, the analytic expressions accurately reproduce the lowest branch in the small-$z_\ell$ regime, within the range of validity of the low-energy expansion.
We emphasize that the low-energy approximation captures only the long-wavelength limit of the lowest surface branch, whereas the full numerical solution contains an infinite tower of normal modes for each angular momentum $\ell$.

We conclude this subsection by highlighting several salient features of the low-energy spectrum.

First, the $\ell=1$ modes correspond to rigid translations of the droplet and therefore remain gapless, $\omega_{1}=0$. 
This reflects the translational invariance of the underlying system and the fact that the droplet is self-bound rather than confined by an external potential. 
For $\ell \geq 2$, the modes are genuine surface oscillations driven by surface tension, with the low-energy frequencies scaling as $\omega_{\ell} \propto \sqrt{\xi}$ for fixed $\ell$.
At large $\ell$, the spectrum asymptotically behaves as
\begin{equation}
 \omega_{\ell} \propto 
 \sqrt{\ell (\ell-1)(\ell+2)} \xrightarrow{\ell \gg 1}
 \ell^{3/2},
\end{equation}
which is consistent with the well-known capillary-wave dispersion $\omega \sim k^{3/2}$ for a planar interface~\cite{Landau:Fluid}.

In contrast, the breathing mode ($\ell=0$) exhibits qualitatively different behavior. 
Its frequency scales as $\omega_{0}\propto\sqrt{3-\xi}$ and softens as $\xi\to3^{-}$, signaling a mechanical marginal stability of the droplet. 
At this point, the restoring effect of the bulk compressibility and the destabilizing tendency associated with surface tension balance each other. 
For $\xi>3$, the breathing mode becomes unstable, indicating the loss of a stable self-bound configuration.

It is also instructive to contrast this result with the fixed-chemical-potential setup.
In that case, the nonlocal term originating from the particle-number constraint is absent.
Nevertheless, the breathing mode can remain stable due to the dynamical bulk response encoded in the boundary action.
Imposing the particle-number constraint introduces an additional nonlocal contribution to the effective action, which further modifies the restoring force of the breathing mode.

\subsection{Canonical quantization and ripplons}
\label{sec:quantization-in-vacuum}

We now quantize the low-energy surface oscillations.
While the previous subsection established the classical normal-mode spectrum,
quantization promotes these modes to elementary excitations of the droplet.
This enables us to construct the corresponding Fock space,
define single- and multi-ripplon states,
and analyze their angular-momentum properties and selection rules.

As shown in Eq.~\eqref{eq:surface-action-low-energy-time} of the previous subsection, the low-energy effective Lagrangian for surface oscillations is given by
\begin{align}
 L_{\eff}
 &= \sum_{\ell=0}^\infty \sum_{m=-\ell}^\ell
 \left[
  \frac{1}{2} B_{\ell} |\dot{\alpha}_{\ell m} (t)|^2 
  - \frac{1}{2} C_{\ell} |\alpha_{\ell m} (t)|^2
 \right] 
 \notag \\
 &= \sum_{\ell=0}^\infty \sum_{m=-\ell}^\ell
 (-1)^m
 \left[
  \frac{1}{2} B_{\ell}  \dot{\alpha}_{\ell -m} \dot{\alpha}_{\ell m}
  - \frac{1}{2} C_{\ell} \alpha_{\ell -m}  \alpha_{\ell m}
 \right] ,
 \label{eq:Lagrangian-ripplon}
\end{align}
where we used the reality condition $\alpha_{\ell m}^\ast(t)=(-1)^m \alpha_{\ell,-m}(t)$ in the second line.

The Lagrangian \eqref{eq:Lagrangian-ripplon} describes a collection of mutually decoupled harmonic oscillators labeled by the angular quantum numbers $\ell$ and $m$.
We can therefore apply the standard canonical quantization procedure to formulate the quantum theory of the low-energy surface oscillations.
The resulting quanta of these surface oscillations are known as ripplons, also referred to as (surface) phonons in nuclear physics.

The canonical quantization of the surface oscillations is accomplished in the standard manner (see, e.g., \cite{Bohr1952,Bohr-Mottelson1953,Bohr-Mottelson1974}).
We first introduce the canonical momentum $\pi_{\ell m}$ conjugate to $\alpha_{\ell m}$ as
\begin{equation}
 \pi_{\ell m}  
 := \frac{\partial L_{\eff}}{\partial \dot{\alpha}_{\ell m}}
 = (-1)^m B_{\ell} \dot{\alpha}_{\ell - m} ,
\end{equation}
which satisfies $\pi_{\ell m}^* = (-1)^m \pi_{\ell -m}$ following from the reality condition.
We then promote these canonical variables to quantum-mechanical operators $\halpha_{\ell m}$ and $\hpi_{\ell m}$, which satisfy the canonical commutation relations
\begin{subequations}
\begin{align}
 \big[\halpha_{\ell m}, \hpi_{\ell' m'}\big] 
 &= \rmi \delta_{\ell\ell'} \delta_{mm'},
 \\
 \big[\halpha_{\ell m}, \halpha_{\ell' m'}\big] 
 &= \big[\hpi_{\ell m}, \hpi_{\ell' m'}\big]  
 = 0.
\end{align}
\end{subequations}
The corresponding Hamiltonian operator is then given by
\begin{align}
 \hH_{\eff}
 &= \sum_{\ell=0}^\infty \sum_{m=-\ell}^{\ell} 
 (-1)^m
 \left[
  \frac{1}{2B_{\ell}} \hpi_{\ell -m} \hpi_{\ell m}
  + \frac{1}{2} C_{\ell} \halpha_{\ell -m} \halpha_{\ell m}
 \right]
 \notag \\
 &= \sum_{\ell=0}^\infty \sum_{m=-\ell}^{\ell} 
 \left[
  \frac{1}{2B_{\ell}} |\hpi_{\ell m}|^2
  + \frac{1}{2} C_{\ell} |\halpha_{\ell m}|^2
 \right],
\end{align}
which shows that each $(\ell,m)$ mode behaves as an independent quantum harmonic oscillator.
We note that the classical reality condition is promoted to the operator identity
\begin{equation}
 \halpha^{\dag}_{\ell m} = (-1)^m \halpha_{\ell,-m}, \quad    
 \hpi^\dag_{\ell m} = (-1)^m \hpi_{\ell -m}.
\end{equation}

We now define the creation and annihilation operators $\hb_{\ell m}^\dag$ and $\hb_{\ell m}$ as
\begin{subequations}
\begin{align}
 \hb_{\ell m}^\dag
 &:= \sqrt{\frac{B_{\ell} \omega_{\ell}}{2}}
 \left( 
  \halpha_{\ell m} - \frac{\rmi}{B_{\ell} \omega_{\ell}} \hpi_{\ell m}^\dag
 \right),
 \\
 \hb_{\ell m}
 &:= \sqrt{\frac{B_{\ell} \omega_{\ell}}{2}}
 \left( 
  \halpha_{\ell m}^\dag + \frac{\rmi}{B_{\ell} \omega_{\ell}} \hpi_{\ell m}
 \right),
\end{align}
\end{subequations}
which satisfies the following commutation relation:
\begin{subequations}
\begin{align}
 \big[\hb_{\ell m}, \hb_{\ell' m'}^\dag\big]
 &= \delta_{\ell\ell'} \delta_{mm'},
 \\
 \big[\hb_{\ell m}, \hb_{\ell' m'} \big]
 &= \big[ \hb_{\ell m}^\dag, \hb_{\ell' m'}^\dag \big] 
 = 0.
\end{align}
\end{subequations}
As usual, the Hamiltonian operator is then expressed as 
\begin{equation}
 \hH_{\eff}
 = \sum_{\ell=0}^\infty \sum_{m=-\ell}^\ell 
 \omega_{\ell}
 \left( \hb_{\ell m}^\dag \hb_{\ell m} + \frac{1}{2} \right) .
\end{equation}
The corresponding energy eigenvalues, labeled by the occupation numbers
$\{ n_{\ell m} \}$, are given by
\begin{equation}
 E \big(\{ n_{\ell m} \}\big)
 = \sum_{\ell=0}^\infty \sum_{m=-\ell}^\ell
 \omega_{\ell}
 \left( n_{\ell m} + \frac{1}{2} \right),
\end{equation}
where we introduced the occupation number $n_{\ell m} = 0,1,2,\cdots$ for each angular quantum number.
We note that the creation/annihilation operators with negative $m$ are not independent degrees of freedom, since the definition implies the constraint $\hb_{\ell,-m} = (-1)^m \hb_{\ell m}^\dag$.
Accordingly, when labeling the multi-ripplon states, it is sufficient to specify occupation numbers for $m \ge 0$.

The operators $\hb_{\ell m}^\dag$ and $\hb_{\ell m}$ allow us to define the ground state $\ket{0}$ by the condition $\hb_{\ell m}\ket{0}=0$, and to construct energy eigenstates by acting the creation operator $\hb_{\ell m}^\dag$ on the ground state.
In particular, the state
\begin{equation}
 \ket{1_{\ell m}} = \hb_{\ell m}^\dag \ket{0},
\end{equation}
which has unit occupation number in the $(\ell,m)$ mode, is an eigenstate of the Hamiltonian with excitation energy $\omega_\ell$ and carries angular momentum quantum numbers $(\ell,m)$.
This state represents a single quantum of surface oscillation, namely a ripplon (surface phonon).
More generally, multi-ripplon states can be constructed straightforwardly by repeatedly acting the creation operators on the ground state.

This formulation allows us to carry over many insights from the study of surface oscillations in nuclear liquid-drop models~\cite{Bohr1952,Bohr-Mottelson1953,Bohr-Mottelson1974}.
One notable example is the selection rules governing the total angular momentum of multi-ripplon states.
For instance, a two-ripplon state composed of identical modes with $\ell = 2$ allows only total angular momentum $L = 0, 2, 4$, while $L = 1, 3$ are forbidden.
These restrictions follow from angular-momentum coupling and Bose symmetry.
They can be exploited to analyze allowed transitions and decay processes involving multi-ripplon excitations.
A systematic investigation of these quantum-mechanical aspects, including detailed spectroscopy and transition rules, is left for future work.

\section{Application to cold atomic droplets}
\label{sec:application-cold-atom}

We now apply the general formalism developed above to a concrete microscopic realization, namely a weakly interacting two-component Bose mixture that forms a self-bound quantum droplet stabilized by beyond-mean-field effects~\cite{Petrov:2015}.
Reviewing the simplified energy density for the two-component Bose mixture in Sec.~\ref{sec:energy-density-2-BECs}, we show how this system can be cast into the form required for our setup in Sec.~\ref{sec:EFT-2-BECs}.
We then specify the bulk parameters of the droplet in Sec.~\ref{sec:bulk-parameters-2-BECs}, leaving the surface tension as the only remaining unknown parameter.

\subsection{Energy density for weakly interacting two-component Bose gases}
\label{sec:energy-density-2-BECs}
For completeness, we briefly review a standard derivation of the bulk equation of state for quantum droplets realized in two-component Bose mixtures, following Ref.~\cite{Petrov:2015}.
At the mean-field level, the energy density of a weakly interacting two-component Bose gas is given by
\begin{align}
 \epsilon_{\mathrm{MF}}(n_{1},n_{2})
 = \frac{\lambda_{11}}{2} n_{1}^{2}
 + \frac{\lambda_{22}}{2} n_{2}^{2}
 + \lambda_{12} n_{1} n_{2}.
\end{align}
Here $n_{1}$ and $n_{2}$ denote the particle number densities, and $\lambda_{ij}$ are the corresponding coupling constants specified below.

In the low-energy sector of interest, fluctuations of the relative particle-number density are stiff and can be eliminated, leaving an effective description in terms of the total density $n:=n_1+n_2$.
This motivates us to introduce the relative particle-number density $\delta n := (n_1 - n_2)/2$ and eliminate it by imposing the stationarity condition for the local energy density $\epsilon_{\mathrm{MF}}(n_1,n_2)$ at fixed $n$:
\begin{align}
 0=\left.\frac{\partial \epsilon_{\mathrm{MF}}}{\partial(\delta n)}\right|_{n}.
\end{align}
Solving this condition, we obtain
\begin{align}
\delta n
=\frac{\lambda_{22}-\lambda_{11}}{2\,(\lambda_{11}+\lambda_{22}-2\lambda_{12})}\,n,
\end{align}
or equivalently, the equilibrium composition
\begin{align}
\frac{n_2}{n_1}=\frac{\lambda_{11}-\lambda_{12}}{\lambda_{22}-\lambda_{12}}.
\end{align}
In particular, near the mean-field instability 
$\lambda_{12}^{2}=\lambda_{11}\lambda_{22}$ with $\lambda_{12}<0$, 
this reduces to the familiar relation 
$n_2/n_1 \simeq \sqrt{\lambda_{11}/\lambda_{22}}$.
Substituting the stationary value of $\delta n$ back into the mean-field energy density yields an effective single-component description  $\epsilon_{\mathrm{MF}}(n)$ expressed solely in terms of the total density $n$,
\begin{equation}
 \epsilon_{\mathrm{MF}} (n)
 = \frac{\lambda_{\mathrm{eff}}}{2} n^2,
\end{equation}
where we introduced the effective coupling constant as
\begin{equation}
\lambda_{\mathrm{eff}}
 := 
 \frac{\lambda_{11} \lambda_{22}-\lambda_{12}^{2}}{\lambda_{11}+\lambda_{22}-2\lambda_{12}}.
\end{equation}

Up to this point, we have focused on the mean-field energy density. 
Beyond the mean-field approximation, quantum fluctuations give rise to the Lee--Huang--Yang (LHY) correction~\cite{Lee:1957a,Lee:1957}.
Including the leading LHY contribution, the effective bulk energy density takes the form
\begin{align}
 \epsilon(n)
 = \frac{\lambda_{\mathrm{eff}}}{2} n^{2}
 + \kappa_{\mathrm{LHY}} 
 n^{5/2}.
 \label{eq:energy-density}
\end{align}
The key ingredient for droplet formation is that the mean-field term can be attractive, $\lambda_{\rm eff}<0$, whereas the leading beyond-mean-field (LHY) contribution is repulsive, $\kappa_{\rm LHY}>0$. 
Their competition stabilizes a self-bound droplet at a finite equilibrium density.
In general, $\kappa_{\mathrm{LHY}}$ depends on microscopic details of the mixture, such as the mass ratio, coupling constants, and the equilibrium composition~\cite{Ancilotto:2018,Cikojevic:2019,Minardi:2019,Ota:2020,Hu:2020,Gurbli:2021}, whose explicit form in a simple setup will be given below.
Thus, extended Gross--Pitaevskii analyses based on Eq.~\eqref{eq:energy-density} indicate that self-bound droplets form with a well-defined interface~\cite{Petrov:2015}.
In regimes where the droplet radius is much larger than the interfacial thickness, our thin-wall effective theory provides an appropriate starting point.

We now specify the form of $\kappa_{\mathrm{LHY}}$. 
To provide an example form of $\kappa_{\mathrm{LHY}}$, we consider a symmetric two-component Bose mixture with equal masses $m_1=m_2 =  m$ and equal intraspecies couplings
\begin{align}
\lambda_{11}=\lambda_{22}
= \lambda = \frac{4\pi}{m}a,
\qquad (a>0),
\end{align}
where $a$ is the $s$-wave 
intraspecies scattering length.
The interspecies coupling is also related to the scattering length $a_{12}$ as $\lambda_{12}=4\pi  a_{12}/m$.
In this symmetric case, minimizing the energy density at fixed $n$ locks the composition to $n_1=n_2=n/2$.
The droplet regime corresponds to attractive interspecies interactions close to the mean-field collapse threshold, which in the symmetric case is $\lambda_{12}\simeq-\lambda$.
We therefore parameterize
\begin{align}
\lambda_{12}=-(1+\delta)\lambda,
\qquad 0<\delta\ll 1,
\label{eq:symmetric_delta}
\end{align}
so that the effective mean-field coupling becomes
\begin{align}
\lambda_{\rm eff}=-\frac{\delta}{2}\lambda<0.
\end{align}
In this setting, the coefficient $\kappa_{\mathrm{LHY}}$ entering Eq.~\eqref{eq:energy-density} is given by~\cite{Petrov:2015}
\begin{align}
 \kappa_{\rm LHY}=\frac{256\sqrt{\pi}}{15}\frac{1}{m}a^{5/2}
 =\frac{8}{15\pi^2}m^{3/2}\lambda^{5/2}.
\label{eq:kappa_symmetric}
\end{align}

\subsection{Effective theory for weakly interacting two-component Bose gases}
\label{sec:EFT-2-BECs}

In this section, we derive the bulk phonon effective Lagrangian based on the equation of state obtained in the previous section.
We begin with the long-wavelength hydrodynamic effective Lagrangian~\cite{Popov2001functional} for a two-component Bose gas,
\begin{align}
\mathcal{L}=
\sum_{a=1,2}\biggl[
    -n_a\partial_t\theta_a
    -\frac{n_a}{2m_a}(\bnab\theta_a)^2
\biggr]
-\epsilon(n)+\mu n,
\label{eq:L_start}
\end{align}
where $\theta_a$ and $n_a$ denote the phase and number density of each component, and $m_a$ are the particle masses.
The energy density $\epsilon(n)$ depends only on the total density $n$, since fluctuations of the relative particle-number density have already been integrated out in the previous section by minimizing the microscopic energy density at fixed $n$.
We also note that the chemical potential $\mu$ is read off from Eq.~\eqref{eq:energy-density} as
\begin{equation}
 \mu = \frac{\partial \epsilon}{\partial n} 
 = \lambda_{\mathrm{eff}} n 
 + \frac{5}{2} \kappa_{\mathrm{LHY}} 
 n^{3/2}.
 \label{eq:mu-n}
\end{equation}

The remaining low-energy dynamics relevant for surface motion are dominated by the in-phase sector, in which the two components share the same superfluid velocity,
\begin{align}
 \bm{v}_1=\bm{v}_2 = \bm{v},
 \quad
 \bm{v}_a:=\frac{\bnab\theta_a}{m_a},
 \label{eq:comoving}
\end{align}
which implies $\bnab\theta_a=m_a\bm{v}$ for $a=1,2$.
To describe the in-phase mode, we introduce the number-density-weighted average mass
\begin{align}
 M:=\frac{m_1n_1+m_2n_2}{n_1+n_2},
\end{align}
and parameterize the phase fields by a single in-phase field $\varphi$ as 
\begin{align}
 \theta_a := \frac{m_a}{M}\,\varphi
 \quad (a=1,2).
\label{eq:theta_inphase}
\end{align}
Note that once the composition is fixed in the bulk to leading order, $n_a=x_a n$ with approximately constant $x_a$, the average mass $M=\sum_a x_a m_a$ can be treated as a constant.
Substituting Eq.~\eqref{eq:theta_inphase} into the phase-kinetic part of Eq.~\eqref{eq:L_start} leads to
\begin{align}
\sum_a
\biggl[
 - n_a \partial_t\theta_a
 - \frac{n_a}{2m_a}(\bnab\theta_a)^2
\biggr]
=
-n \partial_t\varphi-\frac{n}{2M}(\bnab\varphi)^2,
\label{eq:n_psi}
\end{align}
which has the same form as for a single-component superfluid.

As a result, the Lagrangian density for the in-phase sector takes the form
\begin{align}
\Lcal = n X -\epsilon(n)
\label{eq:L_nX_minus_eps}
\end{align}
with the Galilean-invariant combination $X=\mu-\partial_t\varphi-(\bnab\varphi)^2/(2M)$.
This is the phase-space form of the superfluid EFT Lagrangian.
Indeed, the density $n$ appears without derivatives in Eq.~\eqref{eq:L_nX_minus_eps} and can therefore be eliminated by its equation of motion.
After integrating out $n$, the Lagrangian density can be written as $p(X)$, where $p(\mu)$ is the zero-temperature pressure defined by the Legendre transform $p(\mu)=\mu n-\epsilon(n)$.
This reproduces Eq.~\eqref{eq:phonon-Lagrangian}.

\subsection{Bulk parameters in the large-droplet limit}
\label{sec:bulk-parameters-2-BECs}

Although the full functional form of the pressure provides a general effective Lagrangian, the description of fluctuations near the ground state requires only a small set of bulk inputs.
In particular, expanding around the stationary spherical background, the relevant information entering the effective Lagrangian reduces to three quantities evaluated in the ground state: the pressure $\bar{p}$, the number density $\bar{n}=\partial \bar{p}/\partial \mu$, and the compressibility $\chi=\partial \bar{n}/\partial\mu$.
To derive these quantities, we make use of the pressure expressed as a function of $n$
\begin{equation}
 p = \mu n - \epsilon 
 = \frac{1}{2} \lambda_{\mathrm{eff}}n^2
 + \frac{3}{2} \kappa_{\mathrm{LHY}}n^{5/2},
 \label{eq:p-Bose-gas}
\end{equation}
which follows from Eqs.~\eqref{eq:energy-density} and~\eqref{eq:mu-n}.

The ground-state pressure $\bar{p}$ is determined by the Young--Laplace relation~\eqref{eq:Young-Laplace} as $\bar{p}=2\sigma/R_0$.
Given Eq.~\eqref{eq:p-Bose-gas}, one can, in principle, solve the Young--Laplace relation to determine the corresponding number density $\bar{n}$ and the compressibility $\chi$.
To obtain simple analytic expressions, we perform an expansion in powers of $1/R_0$ in the large-droplet limit.
In the large-droplet limit, $\bar p\to 0$ as $R_0\to\infty$, and the bulk density approaches the saturation value $n_0$ determined by $p(n_0) = 0$, yielding
\begin{align}
\bar{n} 
=n_0+O(1/R_0)
= \frac{\lambda^2_{\mathrm{eff}}}{9\kappa^2_{\mathrm{LHY}}}
+O(1/R_{0}),
\label{eq:nbar-2-BECs}
\end{align}
where $O(1/R_{0})$ represents finite-size corrections for a droplet of radius $R_{0}$.

Moreover, differentiating Eq.~\eqref{eq:mu-n} with respect to $n$ immediately gives the inverse compressibility $\chi^{-1}=\partial\mu/\partial n$, yielding
\begin{align}
\chi
=\frac{1}{\lambda_{\mathrm{eff}}+\frac{15}{4}\kappa_{\mathrm{LHY}}\sqrt{\bar{n}}}
=\frac{4}{|\lambda_{\mathrm{eff}}|}+O(1/R_{0}).
\label{eq:chi-2-BECs}
\end{align}
From these results, the sound velocity $c_s$ of bulk phonons is found to be
\begin{align}
c_s=\sqrt{\frac{\bar{n}}{M\chi}}
=\sqrt{\frac{|\lambda_{\mathrm{eff}}|^3}{36M \kappa_{\mathrm{LHY}}^2}}+O(1/R_0).
\end{align}

As we have shown in previous sections, in the thin-wall limit, normal-mode frequencies of the superfluid droplet depend on the microscopic details only through the dimensionless parameter $\xi$ defined in Eq.~\eqref{eq:dimless-parameter} or \eqref{eq:xi-another-expression}.
Since we have specified $\pn$ and $\chi$ in Eqs.~\eqref{eq:nbar-2-BECs}-\eqref{eq:chi-2-BECs}, the only remaining unknown parameter is the surface tension $\sigma$.
Once the surface tension $\sigma$ is fixed by either theoretical input or experimental measurement, our result for the eigenfrequencies provides a universal prediction that can be tested by direct numerical simulations or experiments on the self-bound superfluid droplet.

\section{Concluding remarks}
\label{sec:summary}

In this paper, we investigated the low-energy dynamics of surface oscillations in spherical superfluid droplets.
Using a nonrelativistic superfluid effective field theory (EFT) with a free boundary, we derived an effective action for the surface oscillations.
The resulting effective action~\eqref{eq:action-surface-1} generally contains terms that are nonlocal in time, which arise as remnants of the bulk superfluid phonon contribution.

We then evaluated the normal-mode eigenfrequencies of the surface oscillations for a self-bound superfluid droplet. 
We also analyzed these surface oscillations in the low-energy regime, where the reduced action~\eqref{eq:surface-action-low-energy-time} is characterized by two low-energy parameters: a mass parameter $B_{\ell}$ and an elastic parameter $C_{\ell}$, whose explicit expressions are given in Eq.~\eqref{eq:low-energy-parameters}.
The numerical results for the normal-mode spectrum, as well as the low-energy approximation~\eqref{eq:low-energy-freq}, are shown in Figs.~\ref{fig:plot-zl-l}-\ref{fig:plot-zl-xi}. 
Moreover, we discussed the canonical quantization of these low-energy surface oscillations, resulting in ripplons.
This allows us to formulate a quantum theory of surface oscillations analogous to that underlying the nuclear liquid-drop model.

Finally, we demonstrated that the formalism can be applied to a concrete microscopic system, as illustrated by a weakly interacting two-component Bose gas forming a cold-atom quantum droplet.
In this case, the equation of state can be obtained from the underlying microscopic model, which then fixes all bulk inputs entering the surface EFT.
As a result, the surface spectrum and its low-energy parameters become quantitatively predictable, demonstrating the predictive power and universality of the surface effective theory.

For this application, however, we note that our predictions rely on a controlled low-energy description. Specifically, we treat relative particle-number fluctuations as stiff and eliminate them, focus on the in-phase sector, and employ a thin-wall approximation for the surface.
A natural direction for future work is to relax these assumptions and study interface dynamics in regimes where the out-of-phase mode couples strongly to surface motion and where finite-thickness effects become relevant.
This is particularly important for experimentally realized quantum droplets at finite particle numbers, where the thin-wall approximation and the strict low-energy limit may not be fully satisfied.

Several other interesting directions for future work naturally follow from the present study.
An immediate extension is to generalize the present analysis to a superfluid droplet immersed in another superfluid, namely a superfluid-superfluid interface in the immiscible regime.
In this case, the droplet is in contact with another gapless medium, allowing surface oscillations to undergo radiative damping via the emission of superfluid phonons.
It is also of interest to investigate non-spherical self-bound droplets, for which spherical symmetry is absent and the structure of surface modes becomes correspondingly richer.
Such droplets have been realized, for example, in ultracold dipolar gases~\cite{Kadau:2016,Barbut:2016,Schmitt:2016,Chomaz:2016}.
In these systems, the breaking of spatial rotational symmetry allows mixing between different angular-momentum sectors, leading to the emergence of rotational modes in addition to surface modes.
The coupling between these two types of collective modes, which has been extensively studied in nuclear physics~\cite{Rainwater1950,Bohr1952,Bohr-Mottelson1953,Bohr-Mottelson1974}, would also be interesting to explore in the context of cold-atom systems.

Another important direction is to investigate instabilities of surface modes both in and beyond the linear regime~\cite{Sasaki:2009,Takeuchi:2010}.
Recent work discussed interfacial instabilities in immiscible quantum fluids, including Rayleigh--Taylor--type scenarios~\cite{Geng:2025}.
It would be interesting to formulate such instabilities within the present effective framework, providing a unified description of both stable and unstable surface dynamics.
In addition, the strongly nonlinear regime of large-amplitude deformations, where fission-like shape evolution and eventual fragmentation may occur, as observed in quantum droplets~\cite{Cavicchioli:2025}, remains largely unexplored.
Such studies may also clarify how surface instabilities couple to bulk superfluid phonons once the linear approximation breaks down.

Finally, the present approach may also be extended to a finite droplet in which two distinct superfluid components coexist.
This would provide a useful effective description of surface dynamics in systems such as atomic nuclei composed of proton and neutron superfluids.
The resulting framework may be relevant for understanding pairing correlations in atomic nuclei in vacuum~\cite{Bohr-Mottelson-Pines1958,Belyaev1959,RingSchuck1980,RevModPhys.75.607,BrogliaZelevinsky2013},and for describing nuclear clusters embedded in an $s$-wave neutron superfluid in the inner crust of neutron stars~\cite{Pethick:1995di,Chamel:2008ca,Sedrakian:2018ydt}.
We leave these directions for future work.

\appendix

\acknowledgments
The authors thank Masayuki Matsuo, Kouichi Hagino, and Yoshimasa Hidaka for stimulating discussion. 
M.H. is supported by the Japan Society for the Promotion of Science (JSPS) KAKENHI Grants No. 23K25870, No. 25K01002, and No. 25K07316.
K.F. is supported by JSPS KAKENHI Grant Number 24KJ0062.
This work was partially supported by the RIKEN iTHEMS, the Niigata University Quantum Research Center (NU-Q), and the COREnet project of RCNP at the University of Osaka.

\section{Gauge redundancy and gauge-fixing condition \eqref{eq:fixed-redundancy}}
\label{sec:redundancy}

In this Appendix, we explain the origin of the gauge redundancy associated with the time-dependent chemical potential $\mu(t)$ and justify the gauge-fixing condition \eqref{eq:fixed-redundancy} used in the main text.
The relevant action is given in Eq.~\eqref{eq:action} of the main text.

The key observation is that the action~\eqref{eq:action} is invariant under time-dependent shifts of the phonon field accompanied by a corresponding transformation of the chemical potential:
\begin{equation}
 \begin{cases}
  \varphi (x) \to \varphi (x) + \alpha (t), \\
  \mu (t) \to \mu (t) + \partial_t \alpha (t).
 \end{cases}
 \label{eq:gauge-tr}
\end{equation}
Under the transformation~\eqref{eq:gauge-tr}, the combination $X$ remains invariant.
The last term in Eq.~\eqref{eq:action} changes only by a total time derivative, since the total particle number $N$ is fixed.
Therefore, the action possesses a gauge redundancy under the time-dependent transformation~\eqref{eq:gauge-tr}.

This redundancy allows us to impose an arbitrary condition on the spatially uniform component of the phonon field $\varphi(x)$.
In this paper, we fix this redundancy by choosing the function $\alpha(t)$ such that $\varphi(x)$ satisfies the gauge-fixing condition~\eqref{eq:fixed-redundancy}.
This justifies the use of the gauge-fixing condition~\eqref{eq:fixed-redundancy} in the main text.

\bibliographystyle{utphys}
\bibliography{refs}
\end{document}